\documentclass[5p]{elsarticle}

\usepackage{latexsym}
\usepackage{amsmath}
\usepackage{mathtools}
\usepackage{dsfont}
\usepackage{amsfonts}
\usepackage{graphicx}
\usepackage{boxedminipage}
\usepackage{xcolor} 
\usepackage{subcaption}
\usepackage{bm}
\usepackage[normalem]{ulem}

\usepackage{hyperref}

\usepackage{tikz}
\usetikzlibrary{arrows.meta,calc,decorations.markings,math,arrows.meta}

\DeclareMathOperator*{\real}{Re}
\DeclareMathOperator*{\imag}{Im}
\DeclareMathOperator*{\E}{E}

\newcommand{\PII}{\mathrm{Type-II}}
\newcommand{\PI}{\mathrm{Type-I}}
\newcommand{\LO}{\mathrm{Lomax}}

\providecommand*{\eu}{\ensuremath{\mathrm{e}}}
\providecommand*{\iu}{\ensuremath{\mathrm{i}}}

\renewcommand{\epsilon}{\varepsilon}
\renewcommand{\phi}{\varphi}

\def\Rset{{\mathbb R}}

\graphicspath{{images/}}

\usepackage{etoolbox}
\newtoggle{singlecol}
\togglefalse{singlecol}
\newcommand{\versionflip}[2]{\iftoggle{singlecol}{#1}{#2}}
\newcommand{\imagesize}{\iftoggle{singlecol}{0.5}{0.8}}

\title{On the Sum of Random Samples with Bounded Pareto Distribution}

\begin{document}

\author{Francesco~Grassi\corref{cor1}}
\ead{francesco.grassi@unisalento.it}

\author{Angelo~Coluccia}
\ead{angelo.coluccia@unisalento.it}

\cortext[cor1]{Corresponding author}
\address{Dipartimento di Ingegneria dell'Innovazione, Università del Salento, Via Monteroni, 73100 Lecce, Italy}

\begin{abstract}
Heavy-tailed random samples, as well as their sum or average, are encountered in a number of signal processing applications in radar, communications, finance, and natural sciences. Modeling such data through the Pareto distribution is particularly attractive due to its simple analytical form, but may lead to infinite variance and/or mean, which is not physically plausible: in fact, samples are always bounded in practice, namely because of clipping during the signal acquisition or deliberate censoring or trimming (truncation) at the processing stage. Based on this motivation, the paper derives and analyzes the distribution of the sum of right-censored Pareto Type-II variables, which generalizes the conventional Pareto (Type-I) and Lomax distributions. The distribution of the sum of truncated Pareto is also obtained, and an analytical connection is drawn with the unbounded case. A numerical analysis illustrates the findings, providing insights on several aspects including the intimate mixture structure of the obtained expressions.

\end{abstract}


\maketitle

\section{Introduction}

In the signal processing field, heavy-tailed distributions are important for modeling data found in engineering, finance, natural sciences, and many other fields~\cite{machado2015review}.
In radar signal processing, popular models such as log-normal, Weibull, and compound-Gaussian are adopted to describe the amplitude of non-Gaussian sea clutter returns, especially at low grazing angle and high resolution \cite{Posner}. 
In communications, long tails arise at the network layer e.g. in the modeling of Internet traffic, which does not obey the conventional Poisson assumption~\cite{Internet1,Internet2,Internet3}, as well as at the physical-layer where impulse noise is found~\cite{levey2002statistical}.
In natural sciences, heavy tails are found in different contexts such as hydrology and geology~\cite{mousavi2019stanford}. 

In general, heavy tails arise due to the presence, in the observed distribution of the data, of a small fraction of values more ``extreme'' than the majority of the cases.
Such a behavior can be captured through a power-law relationship, i.e. a function
$
f(x) \propto x^{-\alpha}
$
that (after normalization) leads to the well-known probability density function (pdf) named after Vilfredo Pareto, who introduced it in 1897 to study the distribution of income~\cite{pareto}. This model applies to many types of data and in fact, among the available heavy-tailed pdfs, it remains particularly attractive for signal processing, data analysis, and inference, thanks to its simple analytical form.

The standard (Type I) Pareto distribution is defined for $x\geq \mu$, with $\mu$ a location parameter, and has unbounded upper support. Its behavior is controlled by a shape  parameter $\alpha$, which measures the heaviness of the tail\footnote{Given the underlying power-law functional dependency, the  Pareto cumulative distribution function (CDF) decays as a straight line with slope $-\alpha$ in a doubly-logarithmic plot, see e.g. the dashed line in the inset of Fig.~\ref{fig:comparison}.}: specifically,  $\alpha<2$ yields heavy-tailed variables with infinite second-order moment (variance), while $\alpha<1$ corresponds to even heavier tails for which also the first-order moment (mean) is infinite.
In practice, however, some threshold value can be identified that truncates the tail, in particular when it is known that the variable being modeled is upper bounded due to physical limitations (e.g., saturation in the dynamic range of the signal acquisition circuit~\cite{foi2009clipped}), deliberate replacement of extreme values by a certain fixed value to increase the robustness to outliers (e.g., winsorizing~\cite{hastings1947low}), or their filtering (e.g., trimming~\cite{miao2014additive}).
In this respect, the truncated Pareto distribution~\cite{cohen,inmaculada} can be regarded as a more plausible model, since the heavy-tailed behavior is still captured but finite mean and variance are provided.
Similarly, the right-censored Pareto distribution is more adequate to represent observations that can incur in saturation or clipping effects, hence accumulate at the upper edge of the data range.
Such more physically-plausible Pareto distributions with bounded support have indeed attracted interest in recent years, as numerous evidence shows their suitability to model the statistical behaviour of phenomena arising in radar detection~\cite{mehanaoui2015trimmed, weinberg2018trimmed}, modelling of communication network~\cite{ali2006truncated}, survival analysis~\cite{howlader2002bayesian}, hydrology and atmospheric science \cite{aban2006parameter, gencc2021products}, and natural systems at large~\cite{burroughs2001upper}.

In such contexts, the sum of Pareto variables also naturally arises. 
For instance,  in the radar field, where the Pareto distribution has been recently proposed as a more tractable model for the signal (after square-law detection) in presence of non-Gaussian clutter~\cite{xue2018knowledge},
the sum of Pareto variables is involved in the estimation of the noise power level through cell-averaging CFAR techniques \cite{brennan1968comparison,weinberg2014management}. In communications, the sum of Pareto variables describes the aggregated user requests of network resources (such as connections, calls, data packets) in key performance indicators for mobile cellular systems \cite{coluccia2014robust}, and plays a more general role in the estimation of global parameters in heterogeneous opportunistic wireless sensor networks \cite{coluccia2016robust}.
In seismology, the Pareto distribution can be used to describe the seismic energy released in earthquakes and, consequently, the total seismic energy released by multiple earthquakes as sum of Pareto variables~\cite{zaliapin2005approximating}. In finance, the sum of Pareto random variables has been adopted to describe portfolio's aggregate losses or total catastrophe risk bonds~\cite{goovaerts2005tail,nadarajah2018sums}.
However, while the distribution of the sum of Pareto variables has been investigated in its exact and approximated form \cite{zaliapin2005approximating, blum1970sums, ramsay2008distribution}, the sum of Pareto variables with bounded support has received much less attention.

In this work we derive a convenient expression for the distribution of the sum of right-censored Pareto Type-II variables, which generalizes the conventional Pareto (Type-I) as well as the Lomax distributions (included as special cases of the provided results). An analytical connection is also drawn with the  Laplace-transform based derivation of the unbounded Lomax case in~\cite{ramsay2008distribution}. Furthermore, the distribution of the sum of truncated Pareto variables is obtained. A numerical analysis illustrates the findings for different parameters, providing insights on several aspects including the intimate mixture structure of the obtained pdfs.

The rest of the paper is organized as follows. In Section 2 we recall the concepts of truncated and right-censored distributions, and provide the general expression for the pdf of Pareto Type-II variables with bounded support (either truncated or right-censored). In Section 3 we derive the analytical results, while Section 4 is devoted to the analysis and numerical results. The paper concludes in Section 5.

\section{General Expressions for the Truncated and Right-Censored Pareto Distributions}

Let us consider a set of $n$ i.i.d. random variables $X_i$ with Pareto Type-II distribution having common pdf defined as
\begin{equation}
    f_{X}(x; \mu,\beta,\alpha) = \frac{\alpha\beta^\alpha}{(\beta+x-\mu)^{\alpha+1}}
    \mathds{1}_{\{x \geq \mu\}} \label{eq:pdfX}
\end{equation}
with scale $\beta>0$, shape $\alpha>0$, and location $\mu\in\Rset$, i.e. $X_i\sim P_{\PII}(\mu, \beta, \alpha)$. The indicator function $\mathds{1}_{\{x\in \mathcal{A}\}}$ is $1$ when $x\in \mathcal{A}$, and $0$ otherwise. We denote the sum by $S_n=X_1+X_2+\dots+X_n$ and its pdf by $f_{S_n}$. As mentioned before, the unboundedness of the Pareto distribution might be not plausible in practice. 
Thus, in the following we consider the right-censored random variables $Y_i = \min{(X_i,\tau)}$, where $\tau>\mu$, and $S_n^\tau$ their sum, i.e. $S_n^\tau=Y_1+Y_2+\dots+Y_n$, and study the distribution of $S_n^\tau$. 

We observe that $Y_i$ is a mixed random variable and the common pdf of the $Y_i$'s  can be written as
\begin{equation}
    f_{Y}(y) = (1-p)\delta(y - \tau) + \frac{\alpha\beta^\alpha}{(\beta+y-\mu)^{\alpha+1}}
    \mathds{1}_{\{\mu \leq y < \tau\}}
\end{equation}
where $p= F_{X}(\tau)= 1- (\frac{\beta}{\beta-\mu+\tau})^{\alpha}$ is the value of the CDF\footnote{The CDF of $X$ is obtained by integrating the corresponding pdf \eqref{eq:pdfX}.} of $X$ in $\tau$. 
The pdf of $S_n^{\tau}$ can be obtained by means of the law of total probability as\footnote{In the following, for notation convenience the dummy variable $t$ is used in all pdfs pertaining to the sum of random variables.}
\versionflip{
\begin{equation}
    f_{S_n^{\tau}}(t) =\, (1-p)^{n}\delta\left(t-\tau n\right)
    + \sum_{k=1}^{n} \binom{n}{k} p^k (1-p)^{n-k}f_{T_k^\tau}(t-(n-k)\tau)
    \label{eq:pdf_S_i}
\end{equation}}
{
 \begin{equation}
 \begin{split}
     f_{S_n^{\tau}}(t) =\, &(1-p)^{n}\delta\left(t-\tau n\right)\\
     &+ \sum_{k=1}^{n} \binom{n}{k} p^k (1-p)^{n-k}f_{T_k^\tau}(t-(n-k)\tau)
     \end{split}
     \label{eq:pdf_S_i}
 \end{equation}
}
where $f_{T_k^\tau}$ is the pdf of the sum of $k$ random variables $\widetilde{X}_i$ distributed as $X_i$, but conditioned on $X_i<\tau$, i.e. the sum of $k$ truncated Pareto random variables with pdf
\begin{equation}
    f_{\widetilde{X}}(\tilde{x}) = \frac{1}{p} \frac{\alpha\beta^\alpha}{(\beta+\tilde{x}-\mu)^{\alpha+1}}
    \mathds{1}_{\{\mu \leq \tilde{x} < \tau\}}
\end{equation}
where $p$ (defined as above) acts as normalization factor. 
The equations above illustrate the strict relationship between the sum of right-censored variables and the sum of \mbox{(right-)truncated} variables, and how the pdf of the latter leads to the former. Moreover, it discloses a mixture structure in the continuous part of pdf, composed by shifted versions of $f_{T_k^\tau}$.

\section{Sum of Truncated Pareto Random Variables}

\subsection{Derivation via characteristic function}
To compute  $f_{T_k^\tau}$  we recall that the pdf of the sum of i.i.d. random variables can be obtained as product of their characteristic functions. Moreover, the characteristic function is proportional to the inverse Fourier transform of the pdf (whenever the latter exists). Leveraging these two well-known facts allows us to compute the pdf of $T_k^\tau$ as
\begin{equation}
    f_{T_k^\tau}(t) = \frac {1}{2\pi}\int_{-\infty}^{+\infty} \eu^{-\iu \xi t} \phi(\xi)^k \,\mathrm{d}\xi
    \label{eq:inverse_formula}
\end{equation}
where the characteristic function $\phi(\xi)$ of the right-truncated Pareto Type-II distribution can be computed as
\versionflip{
\begin{align}
      \phi(\xi) & \equiv \phi_{\PII}(\xi) =  \E\left[\eu^{\iu \xi \widetilde{X}}\right]= \frac{1}{p}\int_\mu^{\tau}\eu^{\iu \xi x} \frac{\alpha\beta^\alpha}{(\beta+x-\mu)^{\alpha+1}}\mathrm{d}x\nonumber\\
      & = \frac{\alpha}{p}\eu^{-\iu(\beta-\mu)\xi}(-\iu \beta \xi)^\alpha
      \big(\Gamma(-\alpha,-\iu \beta \xi) - \Gamma(-\alpha,-\iu (\beta+\tau-\mu) \xi)\big)
\end{align}}
{
\begin{align}
\begin{split}
      \phi(\xi) & \equiv \phi_{\PII}(\xi) =  {} \E\left[\eu^{\iu \xi \widetilde{X}}\right] \\
      &= \frac{1}{p}\int_\mu^{\tau}\eu^{\iu \xi x} \frac{\alpha\beta^\alpha}{(\beta+x-\mu)^{\alpha+1}}\mathrm{d}x\nonumber
      \end{split}
      \\
      \begin{split}
      & = \frac{\alpha}{p}\eu^{-\iu(\beta-\mu)\xi}(-\iu \beta \xi)^\alpha
      \big(\Gamma(-\alpha,-\iu \beta \xi) \\ 
      & \quad - \Gamma(-\alpha,-\iu (\beta+\tau-\mu) \xi)\big)
      \end{split}
\end{align}}
and $\Gamma(\cdot,\cdot)$ is the upper incomplete Gamma function.
$\phi(\xi)$ can be particularized for the two cases of truncated Pareto Type-I and Lomax distributions for $\mu=\beta$ and $\mu=0$, respectively:
\begin{equation}
    \phi_{\PI}(\xi) = \frac{\alpha}{p}(-\iu \beta \xi)^\alpha \big(\Gamma(-\alpha,-\iu \beta \xi) - \Gamma(-\alpha,-\iu \tau \xi)\big)\notag
\end{equation}
\versionflip{\begin{equation}
      \phi_{\LO}(\xi) = \frac{\alpha}{p}\eu^{-\iu\beta \xi}(-\iu \beta \xi)^\alpha \big(\Gamma(-\alpha,-\iu \beta\xi) - \Gamma(-\alpha,-\iu (\beta+\tau) \xi)\big).
      \label{eq:lomax}
\end{equation}}
{
\begin{equation}
\begin{split}
      \phi_{\LO}(\xi) = &\frac{\alpha}{p}\eu^{-\iu\beta \xi}(-\iu \beta \xi)^\alpha \big(\Gamma(-\alpha,-\iu \beta\xi)\\
      &-\Gamma(-\alpha,-\iu (\beta+\tau) \xi)\big).
      \label{eq:lomax}
\end{split}
\end{equation}
}

Clearly, for $k=1$ the inverse transform~\eqref{eq:inverse_formula} returns the pdf of $\widetilde{X}$. For $k>1$, due to the oscillatory nature of the integrand function, the integral is difficult to evaluate numerically. This issue can be partially alleviated by observing that, being $\phi(\xi)$ an Hermitian function, the integral over the entire real axis is equal to twice the one over the positive real line, i.e.
\begin{align}
    f_{T_k^\tau}(t) &= \frac{1}{2\pi} \left(\int_0^\infty \eu^{-\iu t \xi} \phi(\xi)^k\mathrm{d}\xi  + \int_{-\infty}^0 \eu^{-\iu t \xi}\phi(\xi)^k \, \mathrm{d}\xi\right)\nonumber\\
    &= \frac{1}{2\pi} \int_0^\infty \eu^{-\iu t \xi} \phi(\xi)^k  +  \eu^{\iu t \xi}\phi(-\xi)^k \, \mathrm{d}\xi\nonumber\\
    &= \frac{1}{\pi} \int_0^\infty \real[\eu^{- \iu t \xi} \phi(\xi)^k ]\, \mathrm{d}\xi \quad k\mu\leq t \leq k\tau.
    \label{eq:pdf_sum_truncated_pareto2}
\end{align}
Equation~\eqref{eq:pdf_sum_truncated_pareto2}, which essentially remains a Fourier transform, is more convenient since the oscillation towards the negative part of the real axis is avoided. 
We further manipulate it as follows 
\versionflip{\begin{align}
    f_{T_k^\tau}(t) &=\!\frac{1}{\pi} \real\left[\int_0^\infty \eu^{- \iu t \xi} \phi(\xi)^k \, \mathrm{d}\xi \right] \nonumber\\
    &=\!\frac{1}{\pi} \real\bigg[\int_0^\infty \eu^{-\iu \left((t+k(\beta-\mu)) \xi +  \frac{\alpha k\pi}{2}\right)} \bigg(\frac{\alpha}{p}(\beta \xi)^{\alpha} \big(\Gamma(-\alpha,-\iu \beta \xi)
\!-
\!\Gamma(-\alpha,-\iu (\beta\!+\!\tau\!-\!\mu) \xi)\big)\bigg)^k\mathrm{d}\xi \bigg]
    \label{eq:levin_integral}
\end{align}}
{
\begin{align}
\begin{split}
    f_{T_k^\tau}(t) &=\!\frac{1}{\pi} \real\left[\int_0^\infty \eu^{- \iu t \xi} \phi(\xi)^k \, \mathrm{d}\xi \right] \nonumber
    \end{split}
    \\
    \begin{split}
    &=\!\frac{1}{\pi} \real\bigg[\int_0^\infty \eu^{-\iu \left((t+k(\beta-\mu)) \xi +  \frac{\alpha k\pi}{2}\right)} \bigg(\frac{\alpha}{p}(\beta \xi)^{\alpha}\\
    & \big(\Gamma(-\alpha,-\iu \beta \xi)
\!-
\!\Gamma(-\alpha,-\iu (\beta\!+\!\tau\!-\!\mu) \xi)\big)\bigg)^k\mathrm{d}\xi \bigg]
    \end{split}
    \label{eq:levin_integral}
\end{align}}
where we explicitly collected the oscillatory components of $\phi(\xi)$ in the Fourier complex exponential kernel.
In doing so, the integral can be efficiently computed by leveraging numerical integration methods for functions expressed as product between an oscillatory part and decaying part~\cite{bailey1994fast,levin1996fast,ooura1999robust}.

The expression for the CDF follows naturally by applying the Gil-Pelaez inversion integral~\cite{wendel1961nonabsolute}, giving
\begin{equation}
    F_{T_k^\tau}(t) = \frac{1}{2} - \frac{1}{\pi} \imag\left[\int_0^\infty \eu^{- \iu t \xi} \frac{\phi(\xi)^k}{\xi} \, \mathrm{d}\xi \right]
\end{equation}
where the same manipulation as in~\eqref{eq:levin_integral} is possible to leverage numerical integration scheme for oscillatory functions.

\begin{figure}
\centering
\begin{tikzpicture}
\draw [-{Latex[scale=1.5]}] (0,-2.5) -- (0,2.5);  
\draw [-{Latex[scale=1.5]}] (-2.5,0) -- (2.5,0);   
\foreach \y in {-1,...,1} {
  \draw (-4pt,\y) -- (4pt,\y) node[pos=0,left] {};
  \draw (\y,-4pt) -- (\y,4pt) node[pos=0,below] {};
}

\draw[thick,black,xshift=2pt,
decoration={ markings, 
      mark=at position 0.2 with {\arrow{latex}}, 
      mark=at position 0.6 with {\arrow{latex}},
      mark=at position 0.8 with {\arrow{latex}}, 
      mark=at position 0.98 with {\arrow{latex}}}, 
      postaction={decorate}]
  (0,2) -- (0,0.25)  node [pos=0, label={[shift={(-0.4,0.2)}]$\iu R$}] {}  node [pos=1, label=left:$\iu \epsilon$] {}  arc (90:-90:.25) -- (0,-2) node [pos=1, label={[shift={(-0.5,-0.7)}]$-\iu R$}] {}  node [pos=0, label=left:$-\iu \epsilon$] {}  ;
\draw[thick,black,xshift=2pt,
decoration={ markings,
      mark=at position 0.2 with {\arrow{latex}}, 
      mark=at position 0.4 with {\arrow{latex}},
      mark=at position 0.6 with {\arrow{latex}}, 
      mark=at position 0.8 with {\arrow{latex}}}, 
      postaction={decorate}]
 (0,-2) arc (-90:-45:2) node [pos=1, label=right:$c-\iu T $] {} node [midway, label=$\mathcal{C}$] {}  --  ({2*sqrt(2)/2},{2*sin(45)}) arc (45:90:2)  node [pos=0, label=right:$c + \iu T $] {}  node [midway, label=below:$\mathcal{C'}$] {};
 \draw[thick,dashed,black,xshift=2pt,
decoration={ markings,
      mark=at position 0.45 with {\arrow{latex}},
      mark=at position 0.6 with {\arrow{latex}}, 
      mark=at position 0.8 with {\arrow{latex}},
      mark=at position 0.9 with {\arrow{latex}}}, 
      postaction={decorate}]
 (0,-2.1) arc (-90:-45:2.1) -- ({2.1*sqrt(2)/2},{2.1*sin(45)}) arc (45:175:2.1) -- (0,{2.1*sin(175)}) arc (-90:-270:{2.1*sin(185)}) -- ({2.1*cos(185)},{2.1*sin(185)}) arc (-175:-90:2.1);
 \end{tikzpicture}
\caption{Contour $\Omega$ (solid line) of complex integral in~\eqref{eq:contour} in comparison with contour $\Omega'$ (dashed line) adopted in \cite{ramsay2008distribution}.}\label{fig:omega}
\end{figure}
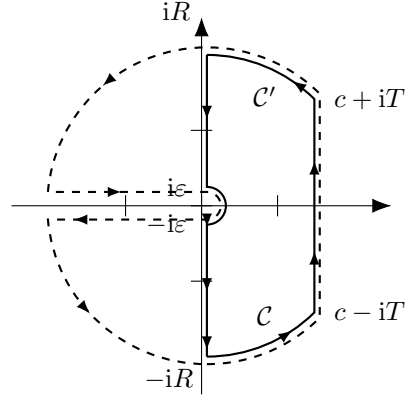

\subsection{Comparison with sum of unbounded Lomax variables}

The result obtained in the previous section makes use of the characteristic function to write the pdf of $S_n^\tau$. Ramsay~\cite{ramsay2008distribution} shows an alternative approach for the sum of Lomax variables, which leverages the fact that the pdf of the sum of i.i.d. random variables is the convolution of their pdfs, and, using the convolution theorem, writes the $k$-th convolutions of the pdfs as their Laplace transform raised to the $k$-th power. Hereby, we will show that such an approach, in case of truncated Pareto variables, leads to the same result we obtained in~\eqref{eq:pdf_sum_truncated_pareto2}.

Let us start considering the following identity
\begin{equation}
    f_{T_k^\tau}(t) = \mathcal{L}^{-1}\left\{\! \left(\mathcal{L}\left\{f_{\widetilde{X}}\right\}(s)\right)^k \right\}(t) \label{eq:inverse_laplace}
\end{equation}
where $\mathcal{L}\{f\}(s)$  is the Laplace transform of $f$ of complex variable $s=\sigma+\iu \omega$ and $\mathcal{L}^{-1}$ is its inverse. The Laplace transform of $f_{\widetilde{X}}$ can be computed as 
\versionflip{
\begin{align}
      f^*(z)  &= \frac{1}{p}\int_\mu^{\tau}\eu^{- s x} \frac{\alpha\beta^\alpha}{(\beta+x-\mu)^{\alpha+1}}\mathrm{d}x\nonumber\\
      &=\frac{\alpha}{p}\eu^{(\beta-\mu)s}(\beta s)^\alpha \big(\Gamma(-\alpha,\beta s)\!-\!\Gamma(-\alpha,(\beta\!+\!\tau\!-\!\mu) s)\big)
\end{align}
}
{
\begin{align}
      &f^*(z)  = \frac{1}{p}\int_\mu^{\tau}\eu^{- s x} \frac{\alpha\beta^\alpha}{(\beta+x-\mu)^{\alpha+1}}\mathrm{d}x\nonumber\\
      &=\frac{\alpha}{p}\eu^{(\beta-\mu)s}(\beta s)^\alpha \big(\Gamma(-\alpha,\beta s)\!-\!\Gamma(-\alpha,(\beta\!+\!\tau\!-\!\mu) s)\big)
\end{align}
}
which is formally identical to the characteristic function~\eqref{eq:lomax} but extended to the entire complex plane.
Letting $c$ and $T$ be positive constants, the inverse Laplace transform is given by the Bromwich integral:
\begin{equation}
    f_{T_k^\tau}(t) = \dfrac{1}{2\pi \iu } \lim_{T\to\infty}\int_{c-\iu T}^{c+\iu T} \eu^{t s} f^*(s)^k \, \mathrm{d}s\label{eq:bromwich}
\end{equation}
which can be solved by contour integration and by means of the residue theorem. When the variables in the sum are not truncated, a valid contour of integration $\Omega'$ (adopted in \cite{ramsay2008distribution}) is the one illustrated in dashed line in Fig.~\ref{fig:omega}. In our case instead, the contour has to be modified to exclude the half-plane containing the negative real axis, where $f^*$ does not converge. The integral is then computed along the contour $\Omega$ (Fig.~\ref{fig:omega}, solid line) as
\begin{equation}
     \int_{\Omega} = \int_{\iu R}^{\iu \epsilon} +  \int_{-\iu\epsilon}^{-\iu R}+\int_{\mathcal{C}}+\int_{\mathcal{C}'}+ \int_{c-\iu T}^{c+\iu T} = 0
     \label{eq:contour}
\end{equation}
where the last equality holds since $f^*$ has no poles inside the contour of integration. For $\epsilon\to 0$, $R\to\infty$, and $T\to\infty$ it is simple to notice that the integrals over the arcs $\mathcal{C}$ and $\mathcal{C}'$ go to zero. Hence
\begin{align*}
      \int_{c-\iu \infty}^{c+\iu \infty}\! \eu^{ts} f^*(s)^k \mathrm{d}s
     \!&=\!\int_{0}^{\iu \infty}\!\eu^{ts} f^*(s)^k \mathrm{d}s-\!\int_{0}^{-\iu \infty}\!\eu^{ts} f^*(s)^k \mathrm{d}s\\
     &=\!\iu\! \int_{0}^{\infty} \eu^{t  \, \iu \omega} f^*(\iu \omega)^k +  \eu^{-t \, \iu \omega} f^*(-\iu \omega)^k \mathrm{d}\omega
\end{align*}
where we replaced $s=\iu \omega$ and $s=-\iu \omega$ in the first and second integral, respectively. Substituting in~\eqref{eq:bromwich} we obtain
\begin{align}
    f_{T_k^\tau}(t) &= \frac{1}{2\pi} \int_0^\infty \left[ \eu^{\iu t \omega} f^*(\iu \omega)^k  + \eu^{-\iu t v} f^*(-\iu \omega)^k \right]\mathrm{d}v \nonumber \\
     &= \frac{1}{\pi} \int_0^\infty \real[\eu^{\iu t \omega} f^*(\iu \omega)^k ] \mathrm{d}\omega \quad k\mu\leq t \leq k\tau.
     \label{eq:pdf_laplace_method}
\end{align}
where we noticed that the two terms in the first equation are complex conjugate. It is immediately evident that~\eqref{eq:pdf_sum_truncated_pareto2} and~\eqref{eq:pdf_laplace_method} are formally equivalent.

\section{Analysis and Numerical Results}

\begin{figure}
    \centering
    \includegraphics[width=\imagesize\columnwidth]{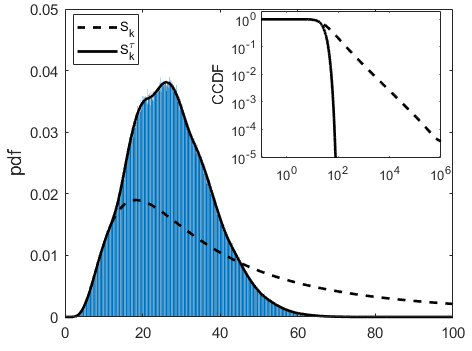}
    \caption{Pdf and CCDF of $S_n$ and $S_n^\tau$ for $n=10$, $\tau=10$, and $X_i\sim \mathrm{Lomax}(1,0.9)$. Normalized histogram of $S_n^\tau$ is obtained for the same parameters via Monte Carlo simulations.}
    \label{fig:comparison}
\end{figure}
\begin{figure}
    \centering
    \includegraphics[width=\imagesize\columnwidth]{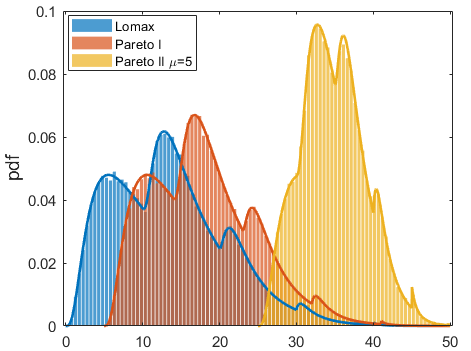}
    \caption{Pdf of $S_n^\tau$ for $n=5$, $\tau=10$, and $X_i\sim\mathrm{Lomax}(1,0.9)$, $X_i\sim P_{\PI}(1,0.9)$, and $X_i\sim P_{\PII}(5, 1, 0.9)$. Corresponding histograms of $S_n^\tau$ for the same parameters are obtained via Monte Carlo simulations.}
    \label{fig:pareto2}
\end{figure}
\begin{figure}
    \centering
    \includegraphics[width=\imagesize\columnwidth]{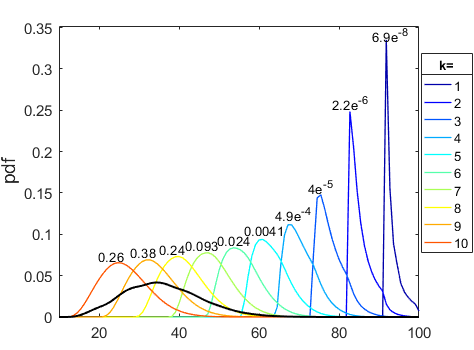}
    \caption{Continuous part of the pdf of $S_n^{\tau}$ (black line) and pdfs of the sum of truncated Pareto $T_k^\tau$ for $k=1,\dots, 10$ with correspondent weights.}
    \label{fig:mixture}
\end{figure}

\begin{figure}
    \centering
    \includegraphics[width=\imagesize\columnwidth]{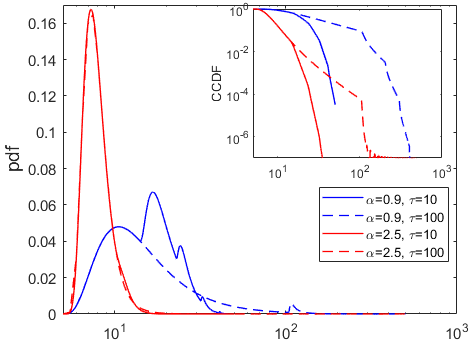}
    \caption{Pdf and CCDF of $S_n^{\tau}$ for $n=5$ and different values of $\alpha$ and $\tau$.}
    \label{fig:parameters}
\end{figure}

In the following, without loss of generality, we will assume the scale parameter to be $\beta=1$. Moreover, when dealing with right-censored random variables, we will only discuss the continuous part of the distribution, since the discrete part, besides being negligible in many common cases, is always a discrete Dirac delta function localized at the right extreme of the pdf support hence is not worth further investigation.

We start by comparing the obtained distribution with the one characterizing the sum of unbounded Lomax random variables described in~\cite{ramsay2008distribution}. Fig.~\ref{fig:comparison} shows the pdfs of the sum of $n$ right-censored Lomax (solid) and the sum of $n$ unbounded Lomax random variables (dashed), with $n=10$, $\alpha=0.9$, and censoring threshold $\tau=10$. We can observe the effects of the censoring on the support of the distribution of $S_n^\tau$, which is limited in the range $[0, n\tau]$, whereas the distribution $f_{S_n}$ exhibits a much heavier tail (and unbounded support). The behaviour of the tails of these distributions is much more clear when looking at the Complementary CDF (CCDF) in the log-log axis, shown as inset in the same figure. This reveals the typical linear negative slope for ${S_n}$, while $S_n^\tau$ density decays with a double exponential rate when approaching the upper limit of its support.
The histogram obtained via Monte Carlo simulation of $S_n^\tau$ with same parameters is also reported in the figure for comparison, showing an excellent match between the theoretical expression and the numerical experiments. 

After investigating the effect of the truncation on the distribution support and tail behaviour, we illustrate how the shape of $f_{S_n^\tau}$ changes for different type of Pareto distributions. In particular, we consider the behaviour of the distribution for $n=5$, $\tau=10$, $\alpha=0.9$ and for different values of the location parameter $\mu$. Fig.~\ref{fig:pareto2} shows the theoretical distribution for $\mu=0$ (Lomax), $\mu=\beta=1$ (Pareto Type-I), and $\mu=5$ (Pareto Type-II) together with the corresponding histograms obtained via Monte Carlo simulations. We observe that all the resulting densities exhibit multimodal profiles, supporting the interpretation of the pdf as a mixture of different components. As expected, changing the location parameter $\mu$ makes the lower limit of the distribution support equal to $n\mu$. More interestingly, we notice that, since the upper limit is fixed at $n\tau$, changing $\mu$ deforms the distribution profile in a non-trivial way, modifying size and height of the modes.

Fig.~\ref{fig:mixture} discloses the mixture structure of the continuous part of $f_{S_n^\tau}$, for $n=10$ and $\tau=10$. In particular, each colored line in the figure represents the pdf of the sum of $k$ truncated Pareto Type-I, with $\alpha=0.9$ and $\beta=\mu=1$. Although in principle the support of each pdf $f_{T_k^\tau}$ is $[k\mu, k\tau]$, we observe that they are also shifted by $(n-k)\tau$, in agreement with~\eqref{eq:pdf_S_i}. Additionally, the contribution of each component to the mixture is weighted by the corresponding value ${\binom{n}{k} p^k (1-p)^{n-k}}$ reported on top of each line, and the resulting weighted sum is shown as solid black line in the same figure. Comparing this pdf and the weights, one can clearly realize that only the last three components ($k\geq 8$) contribute significantly to the pdf.

Finally, we explore the role played by the different parameters in shaping the distribution of $S_n^\tau$. Fig.~\ref{fig:parameters} shows the pdf and the CCDF of the sum of $n=5$ Pareto Type-I random variables for different combinations of the parameters $\tau$ and $\alpha$. First, we can notice that for $\alpha=2.5$ the value of $\tau$ does not impact significantly onto the shape of the pdf. This is expected, since the pdfs constituting the mixture decay so rapidly that the truncation effect is substantially negligible. On the other hand, for $\alpha=0.9$, the value of $\tau$ makes a profound impact, changing the shape of the pdf from a multimodal profile with peaks of different sizes and heights for $\tau=10$, to one with a single dominant mode and very low peaks in the higher part of the support for $\tau=100$. These secondary modes can be explained by noticing that the $k$-th components of the mixture for small $k$ have associated weights which are not negligible when $\tau$ and $\alpha$ are respectively large and small enough. Hence, in general, different combinations of the parameters $\alpha$ and $\tau$ may change dramatically the distribution, from a mild single mode to a more wild multimodal density. The CCDF in the inset of same figure illustrates how the tail of the distribution behaves for the same set of parameters. Here it is even more clear that, for $\alpha=2.5$, the effect of the truncation is hidden by the rapid decay of the power law. In this case and for $\tau=10$, the CCDF for $\tau=10$ reaches values below $10^{-6}$ much earlier than the corresponding curve obtained for $\alpha=0.9$, which instead reaches around $10^{-4}$ only at the end of its support.

\section{Conclusion}
We have derived and studied a convenient expression for the distribution of the sum of right-censored Pareto Type-II samples, which includes the conventional Pareto (Type-I) as well as the Lomax cases. The distribution of the sum of truncated Pareto variables has been also obtained, and an analytical connection has been drawn with the  Laplace-transform based derivation of the unbounded Lomax. A numerical analysis has highlighted relevant behaviors and provided insights on several aspects, including the intimate mixture structure of the obtained pdfs and the consequent possible multi-modality.

\bibliographystyle{elsarticle-num}
\bibliography{biblio.bib}

\end{document}